\documentclass[manuscript,authorversion]{acmart} 

\usepackage{makecell}
\usepackage{multirow}
\usepackage{longtable}

\AtBeginDocument{%
  \providecommand\BibTeX{{%
    \normalfont B\kern-0.5em{\scshape i\kern-0.25em b}\kern-0.8em\TeX}}}

\begin{document}

\title{People's Perceptions Toward Bias and Related Concepts in Large Language Models: A Systematic Review}








 


\author{Lu Wang}
\email{lw823@drexel.edu}
\affiliation{%
  \institution{College of Computing \& Informatics, Drexel University}
  \streetaddress{3675 Market Street}
  \city{Philadelphia}
  \state{Pennsylvania}
  \country{USA}
  \postcode{19104}
}

\author{Max Song}
\email{ms5526@drexel.edu}
\affiliation{%
  \institution{College of Computing \& Informatics, Drexel University}
  \streetaddress{3675 Market Street}
  \city{Philadelphia}
  \state{Pennsylvania}
  \country{USA}
  \postcode{19104}
}

\author{Rezvaneh Rezapour}
\email{sr3563@drexel.edu}
\affiliation{%
  \institution{College of Computing \& Informatics, Drexel University}
  \streetaddress{3675 Market Street}
  \city{Philadelphia}
  \state{Pennsylvania}
  \country{USA}
  \postcode{19104}
}

\author{Bum Chul Kwon}
\email{bumchul.kwon@us.ibm.com}
\affiliation{%
  \institution{IBM Research}
  \city{Cambridge}
  \state{Massachusetts}
  \country{USA}
  \postcode{02142}
}

\author{Jina Huh-Yoo}
\email{jh3767@drexel.edu}
\affiliation{%
  \institution{College of Computing \& Informatics, Drexel University}
  \streetaddress{3675 Market Street}
  \city{Philadelphia}
  \state{Pennsylvania}
  \country{USA}
  \postcode{19104}
}

\begin{abstract}
Large language models (LLMs) have brought breakthroughs in tasks including translation, summarization, information retrieval, and language generation, gaining growing interest in the CHI community. Meanwhile, the literature shows researchers' controversial perceptions about the efficacy, ethics, and intellectual abilities of LLMs. However, we do not know how people perceive LLMs that are pervasive in everyday tools, specifically regarding their experience with LLMs around bias, stereotypes, social norms, or safety. In this study, we conducted a systematic review to understand what empirical insights papers have gathered about people's perceptions toward LLMs. From a total of 231 retrieved papers, we full-text reviewed 15 papers that recruited human evaluators to assess their experiences with LLMs. We report different biases and related concepts investigated by these studies, four broader LLM application areas, the evaluators' perceptions toward LLMs' performances including advantages, biases, and conflicting perceptions, factors influencing these perceptions, and concerns about LLM applications.   
\end{abstract}





\received{20 February 2007}
\received[revised]{12 March 2009}
\received[accepted]{5 June 2009}

\maketitle

\section{Introduction}
Large language models (LLMs), which utilize deep neural networks to learn the relationships between words in natural language and are trained on vast datasets comprising billions of documents sourced from the Internet~\cite{thirunavukarasu2023large, wu2022ai, wei2023overview}, have brought breakthroughs in various tasks, including translation, summarization, information retrieval, and conversational interactions \cite{naveed2023comprehensive}. Microsoft launched the Bing search engine running on an OpenAI LLM customized specifically for search \cite{Bing}. There is growing interest in the CHI community regarding different strategies to prompt LLMs \cite{wu2022ai, dang2023choice}, interaction designs for mobile user interfaces \cite{wang2023enabling}, and applications in various areas such as writing \cite{petridis2023anglekindling, dang2023choice}, coding \cite{liu2023wants}, supporting public health interventions \cite{jo2023understanding}, supporting augmentative and alternative communication \cite{valencia2023less}, and supporting synthetic HCI research \cite{hamalainen2023evaluating}. 

However, LLMs have been documented to inherit social bias from training data \cite{zhuo2023red, ferrara2023should}, leading to unjust treatment of marginalized communities including the unjust association of Muslims with violence \cite{abid2021large}, as well as an excessive representation of gun violence, homelessness, and drug addiction in discussions concerning mental illness \cite{hutchinson2020social}. While researchers are actively working on developing evaluations \cite{zhang2023chatgpt, zhang2023wider, steed2022upstream} to assess the biases in LLMs, some biases are believed inevitable due to the inherent nature of language and cultural norms \cite{ferrara2023should}. The complexity of sociocultural factors complicates the investigation, the integration of diverse tasks within a single model adds to the challenges faced by LLMs, and the disagreements of individual user opinions and preferences contribute to the complexity of the issue. Researchers also assert that creating a perfectly safe model requires models to deeply understand languages until Artificial Intelligence (AI) itself becomes a reality \cite{xu2021bot}.

There are controversial conversations among researchers about the efficacy, ethics, and intellectual abilities of LLMs \cite{michael2022nlp, cabrero2023perceived}. Various research methodologies employ human evaluation to gain a deeper understanding of LLMs and assess their performance more effectively. Given the extensive range of applications for LLMs, it is crucial to gain insight into how individuals perceive LLMs and what kind of biases they might consider significant. Specifically, we will investigate the following research questions:

\begin{itemize}
\item {RQ1}: In studies examining people's perceptions of LLMs, what specific biases did researchers explore?
\item {RQ2}: In which settings or contexts were LLMs applied in these studies?
\item {RQ3}: What were the findings regarding people's perceptions of LLMs as explored in these studies?
\end{itemize}

Through a systematic review, this study informs the different biases of LLMs studied, the existing applications of LLMs evaluated from perspectives of human evaluators, dimensions of people's perceptions of LLM investigated by researchers, and the gaps as well as the opportunities for future research. Developers and designers will benefit from these findings for future development and applications of user-centered LLMs. The insights could also be generalized to human-centered AI design considering the bias. 

\section{Related Work}

\subsection{Natural Language Processing and LLMs} 

Since the introduction of LLMs in Natural Language Processing (NLP), the AI space has become increasingly popular due to its capability of successfully ``understanding'' and generating natural language \cite{zhuo2023red, hadi2023large, ferrara2023should}. Paying homage to the early development of language models, recurrent neural networks, and transformer models, LLMs such as ChatGPT and Google Bard have shown significant success in various language-related tasks such as translation \cite{karpinska2023large}, text generation \cite{lancaster2023artificial}, summarization\cite{zhang2023benchmarking}, question answering \cite{tan2023evaluation}, and sentiment analysis\cite{ferrara2023should, hadi2023large}. 

LLMs boast a diverse array of applications across two primary scenarios: creative generation and decision-making \cite{zhuo2023red}. In creative generation, LLMs are employed to produce innovative and imaginative content, encompassing tasks like crafting a narrative, composing poetry, or scripting dialogue for a film. For decision-making, LLMs are used to make informed decisions based on natural language instruction, which can be observed in tasks on sentiment analysis, text classification, and question answering \cite{zhuo2023red, hadi2023large}.

Furthermore, these LLM applications extend across a variety of different fields, like education (e.g., students are engaging with their course material in entirely new ways) \cite{su2023unlocking, haleem2022era, crompton2023artificial}, finance (e.g., LLMs designed specifically for finance, like BloombergGPT \cite{wu2023bloomberggpt}, have revolutionized financial NLP tasks like risk assessment, algorithmic trading, market prediction and financial reporting \cite{hadi2023large}), engineering (e.g., especially in software engineering, LLMs assist with code generation, debugging, software testing, documentation generation, and collaboration \cite{fraiwan2023review, hadi2023large}), and healthcare (e.g., LLMs have been successfully employed in medical education, radiologic decision-making, clinical genetics, and patient care \cite{kung2023performance, sallam2023chatgpt, gilson2023does}).

\subsection{HCI and LLMs}
There is growing interest in the HCI community regarding different aspects of LLMs, including prompting \cite{wu2022ai, dang2023choice,zamfirescu2023johnny}, interaction designs \cite{wang2023enabling}, applications for various tasks such as coding \cite{liu2023wants, vaithilingam2022expectation} and research \cite{hamalainen2023evaluating}, and supporting different groups in mental health support \cite{jo2023understanding} and communications \cite{valencia2023less}. 

Prompt engineering is a growing interest within the CHI community, as many related works show that para-phrasing prompts can lead to better model outputs \cite{dang2023choice}. For example, chaining prompts for LLMs, a process of breaking up complex tasks into smaller steps to be independently run while the output of one or more steps is used as input for the next, raises the bar of possibility for prompt engineering \cite{wu2022ai}. Besides, by using a set of prompting techniques designed to adapt LLMs to mobile User Interfaces (UIs), the HCI community is trying to make advancements with machine learning, bridging the gap in NLP and graphical UIs (e.g., web UIs and popular systems like iOS and macOS) to enable conversational interaction \cite{wang2023enabling}. However, researchers found that non-AI experts met challenges in prompt engineering in terms of generating prompts, evaluating prompts, and explaining prompts' effects \cite{zamfirescu2023johnny}.

Advancements in LLMs provided the ability to assist users in different tasks. For example, LLMs can automatically generate code in multifarious programming languages, like Python, researchers have been propelled further to study the usability of LLM-based code generation tools, like GitHub Copilot, for real-world programming tasks \cite{vaithilingam2022expectation}. Besides, data collection in HCI research can become a hindrance. Since interviews and questionnaires dominate this area, collecting data turns into a slow and arduous process. In essence, LLMs can be viewed as a newfound search engine with the capability of prompt engineering to query for information in many different ways (e.g., querying in the form of a narrative) \cite{hamalainen2023evaluating}. In the bigger picture, LLMs have the potential to expand computational user modeling and simulation to increasingly new levels. However, their usefulness relies heavily upon the validity of the generated data \cite{hamalainen2023evaluating}.

Moreover, advancements in LLMs enhancing the quality of open-ended conversations with chatbots have alternatively fueled other researchers to study the potential of LLM-driven chatbots, like CareCall, to support public health interventions by monitoring populations at scale through empathetic interactions in real-world settings \cite{jo2023understanding}. Furthermore, in the mental health domain, researchers have shown the potential of LLM's success in therapeutic settings. Teachable self-help techniques, such as mindfulness, can reduce anxiety and improve mental well-being outcomes. Thus, HCI researchers are also exploring is use of LLM's for improving the awareness of mindfulness \cite{kumar2023exploring}. Individuals utilizing augmentative and alternative communication (AAC) devices may encounter challenges in real-time communication due to the time required for composing messages. AI technologies, like LLMs, offer a potential avenue to assist AAC users by enhancing the quality and diversity of text suggestions. Researchers have found that these technologies will radically change how users interact with AAC devices as users transition from typing their own phrases to prompting and selecting AI-generated phrases--highlighting that there are both positive and negative consequences \cite{valencia2023less}.

\subsection{Bias of LLMs and Challenges in Mitigating Bias in LLMs}
While LLMs are rapidly transforming the way people communicate, create, and work \cite{hacker2023regulating}, significant advancements made by LLMs give rise to emerging ethical concerns like potential biases and fairness, reliability, and toxicity \cite{zhuo2023red}. Different types of biases can emerge when using LLMs, which usually occur due to either biases from the original training data or biases from different models \cite{bender2021dangers, ferrara2023should}. 

Bias is a ubiquitous challenge across various domains, albeit with nuanced differences in its definitions. In statistics, bias refers to ``the error that is introduced by approximating a real-life problem, which may be extremely complicated, by a much simpler model''~\cite{james2013introduction}. In cognition, bias refers to the systematic errors of judgments and choices against rationality, recurring predictably in particular circumstances~\cite{daniel2017thinking}. In sociology, bias is referred to as ``prejudice in favor or against a person, group or thing that is considered to be unfair''~\cite{garrido2021survey, allport1954nature, skinner2017catching}. Applying these concepts to LLMs, biases in LLM can be understood as the inaccuracies or limitations LLMs encounter while learning the relationships between words in natural language---in which the LLM might not fully capture the diversity, subtlety, or complexity of natural language as it is used in real-world contexts~\cite{cao2023assessing, ferrara2023should}---or deviations in human-LLM communication that do not align with rational or unbiased language use~\cite{bender2021dangers}. It may very well also be the exhibitions of sociologically biased outputs, demonstrating unfair or prejudiced language patterns towards certain groups or topics, that has recently been garnering increased attention from researchers~\cite{zhuo2023red, abid2021large, hutchinson2020social, kirk2021bias, mcgee2023chat}. For example, predicted jobs by GPT-2 are less diverse and more stereotypical for women than for men \cite{kirk2021bias}. English prompts of ChatGPT reduce the variance in model responses and flat out cultural differences and biasing them towards American culture \cite{cao2023assessing}. ChatGPT seemed to create positive Irish Limericks for liberal politicians and negative Limericks for conservative politicians \cite{mcgee2023chat}. These biases would bring unintended negative impacts if people apply LLMs without any awareness of such issues. In this paper, we define bias as any deviation from the expected performance, regardless of whether the causes are objective or subjective, and irrespective of inaccuracies or limitations.

To detect and measure biases in LLMs, researchers have developed multiple evaluations to assess them in LLMs \cite{zhang2023chatgpt, zhang2023wider, steed2022upstream}. For example, researchers tracked how bias transferred from pre-training to tasks after fine-tuning and encouraged practitioners to focus more on dataset quality and context-specific harms \cite{steed2022upstream}. Some researchers proposed benchmarks comprising carefully crafted metrics with datasets \cite{zhang2023chatgpt} or multi-layer networks with tasks and samples \cite{zhang2023wider} to evaluate biases in LLMs. However, some biases are believed inevitable due to inherent biases in language, the ambiguity of cultural norms, the subjectivity of fairness, and continuously evolving language and culture \cite{ferrara2023should}. Completely eliminating bias from LLMs is a complex, challenging, and ongoing task, requiring developers, researchers, and stakeholders to continue working on reducing bias in LLMs, collaborate with diverse communities, and engage in ongoing evaluations and mitigation methods \cite{ferrara2023should}. Some researchers also assert that we can only create a perfectly safe model until models can deeply understand languages and this is an AI-complete problem, implying that the difficulty of these computational problems is equivalent to making computers as intelligent as people \cite{xu2021bot}.

Given the challenges of mitigating biases in LLMs, it is imperative that designers and users exert efforts to ensure the safe use of LLM applications and to mitigate potential risks unaware by users, which also requires a clear understanding of the people's perceptions of LLMs.
 
\subsection{People's Perceptions of LLMs} 


Generally, NLP tasks, with a specific goal, favor automatic evaluations due to their cost-effectiveness and efficiency, which are essential for automatic benchmarking and fine-tuning of algorithms \cite{novikova2017we}. Human evaluations are employed in certain tasks where automatic evaluation may be unsuitable, such as text generation, e.g., ROUGE score for abstractive summarization \cite{lin2004rouge}, where the quality of generated content can extend beyond conventional (ground-truth) responses \cite{chang2023survey,rezapour2022makes}. Various measurements were applied in human evaluations to understand people's perceptions such as quality \cite{novikova2017we}, accuracy \cite{chang2023survey}, informativeness \cite{novikova2017we}, naturalness \cite{novikova2017we}. Different tasks require different dimensions of perceptions. For example, researchers analyzed 97 style transfer papers in terms of three aspects: style transfer, meaning preservation, and fluency, and discovered that human evaluation protocols lack specificity and standardization \cite{briakou2021review}. Through 165 papers on natural language generation, researchers discovered that there were more than 200 different terms used to evaluate aspects of quality, indicating a pervasive lack of clarity, extreme diversity in approaches for human evaluations, and the urgent need for standard methods and terminology \cite{howcroft2020twenty}. 

A meta-survey of the NLP community revealed that NLP researchers are split almost exactly in half on questions about whether language models understand language, whether the linguistic structure is necessary, and whether expert inductive biases are necessary \cite{michael2022nlp}. This survey also uncovered false sociological beliefs where the community’s predictions on the distributions of the answers don't match reality \cite{michael2022nlp}, indicating the misunderstanding of what others think. The disagreements within the NLP community could slow down communication and lead to wasted effort, missed opportunities, and needless fights \cite{michael2022nlp}. There are differences among people with different expertise in AI, linguistic-related fields, and other fields in terms of adoption rates and willingness to use and trust LLMs \cite{cabrero2023perceived}. People with expertise in AI were more aware of the limitations of natural language generation and specific ethical concerns, particularly regarding privacy and explainability \cite{cabrero2023perceived}, while people with expertise in linguistics, translators, interpreters, or related areas were more cautious about using natural language generation tools and expressed concerns about their impacts on daily life \cite{cabrero2023perceived}.

Considering the different biases of LLMs with such a variety in people's perceptions, it is imperative to systematically investigate people's perceptions toward LLMs.
Understanding the potential factors behind these perceptions can also inform researchers in improving the user experience of LLMs, given the proliferation of LLMs in people's everyday devices.



\section{METHOD}
\begin{figure}[htbp]
  \centering
  \includegraphics[width=0.6\linewidth]{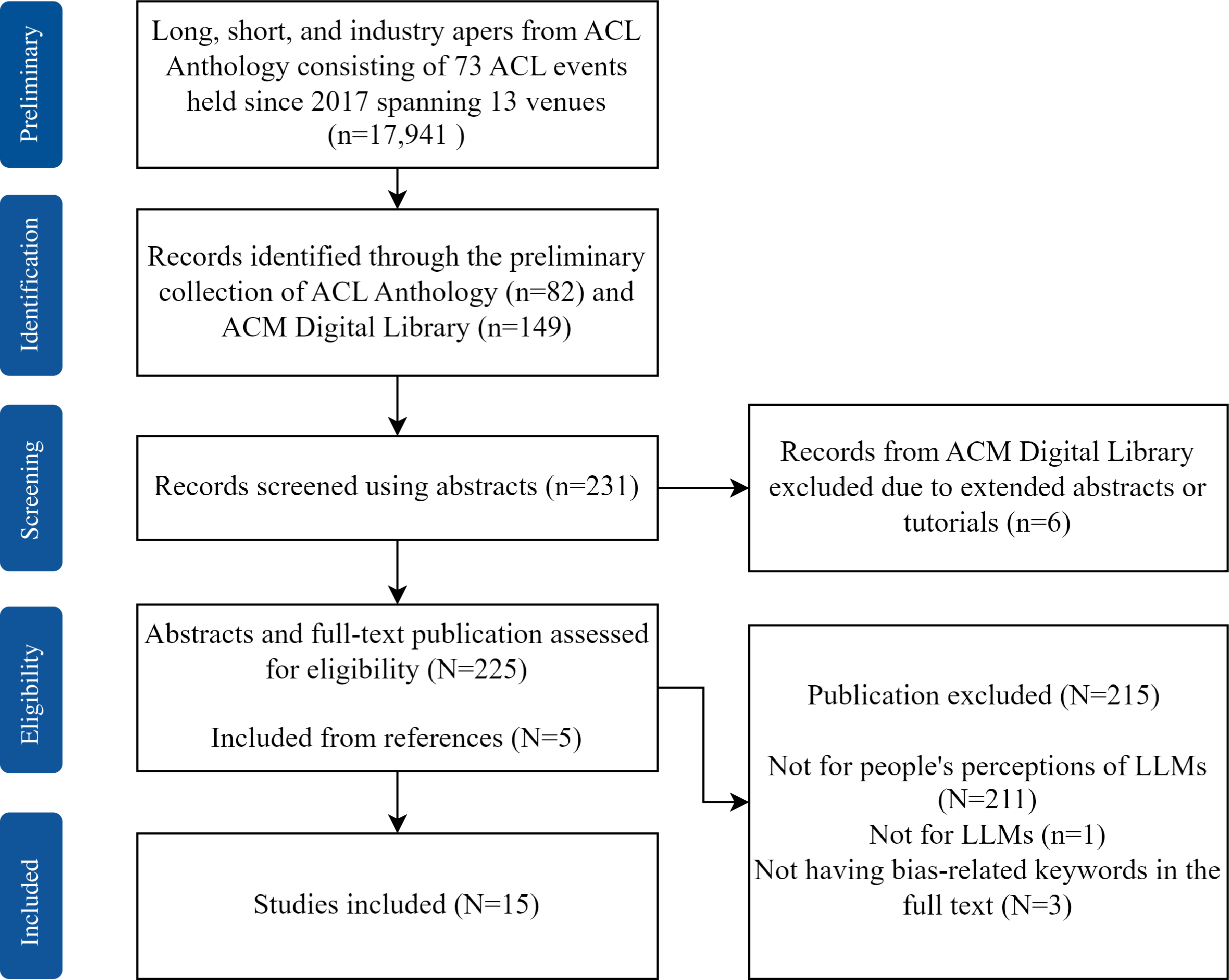}
  \caption{Diagram showing the adapted PRISMA of the review.}
  \Description{This figure shows the PRISMA diagram of the paper selection process including preliminary collections from ACL Anthology, literature identification, screening, eligibility evaluation, and included papers for full-text review.}
  \label{fig:PRISMA}
\end{figure}

We used the Preferred Reporting Items for Systematic Reviews and Meta-Analyses (PRISMA) \cite{page2021prisma} to illustrate our review process (See Fig. \ref{fig:PRISMA}). Following, we introduce the review process in three parts: 1) literature search and screening, 2) eligibility evaluation and backward snowballing, and 3) data extraction and analysis.

\subsection{Literature Search and Screening}
Below, we describe the data source, search query, and filtering and review process.

\subsubsection{Data Source}
We used ACM Digital Library and ACL Anthology databases for the sources of literature collection. 
As a comprehensive bibliographic database focused exclusively on the field of computing, \href{https://dl.acm.org/} {ACM Digital Library} provides access to more than 3 million publications in computing \cite{ACM}. Sponsored by the Association for Computational Linguistics, \href{https://aclanthology.org/} {ACL Anthology} is a digital archive of conference and journal papers in natural language processing and computational linguistics, representing the NLP community's most up-to-date and established freely accessible research repository \cite{bird2008acl}. Since ACL Anthology doesn't provide advanced search functions, \ref{appendix:a} provides details of how we identified papers from ACL Anthology.

\subsubsection{Search Query}
On August 7th, 2023, we conducted a query search on the abstract from ACM Digital Library and ACL Anthology. The search query combined three components using ``AND'' operand between them and ``OR'' within each component: 1) keywords related to people's perceptions: {\itshape experience, perspective*, perception*, acceptability, acceptance*,  reaction*, response*, trust*, usability, accountability, transparenc*, participant*, ``human subject*'', "user stud*", speaker*, listener*}, 2) keywords related to biases: {\itshape bias*, ``social harm*'', stereotype*, stigma*, fair*, norm*, ethic*, safet* }, and 3) keywords related to LLMs: {\itshape ``language model*'', GPT, bert, ``language generation*'', ``contextualized representation*'', ``semantic representation*'', ``generative AI''}. The star sign ``*'' allows matching any number of characters. The quote sign, ``'', allows us to search phrases as a whole so that we can get more relevant results. 
We generated these keywords through brainstorming \cite{rawlinson2017creative} within the research group, and we acknowledge that this list is not exhaustive. In addition, considering LLMs frequently rely on transformers, a neural network architecture introduced by Vaswani et al. in 2017 \cite{vaswani2017attention}, we set the publication date as starting from 2017. 

The search on publications from January 1st, 2017 to August 7th, 2023 resulted in 149 records from the ACM Digital Library and 82 papers from ACL Anthology. Six records from ACM Digital Library were excluded since they were extended abstracts or tutorials. In total, we had 225 papers remaining for eligibility evaluation. 

\subsection{Eligibility Evaluation and Backward Snowballing}

The research group held regular meetings to examine the screened papers, determine the inclusion criteria, and validate the criteria. Studies satisfying all the criteria were included: 1) empirical research involving human evaluations, 2) having relevant statements by using the keywords, bias*, ``social harm*'', stereotype*, stigma*, fair*, norm*, ethic*, and safet*, in the full text, indicating what kind of biases the researchers investigated, and 3) reporting people's direct perceptions of LLMs.

Two authors conducted a full-text review of the 225 papers based on the inclusion criteria, and ten papers remained. Examples of studies that did not meet inclusion criteria that were excluded were: 1) solely relying on automatic evaluations, 2) only using the keywords rhetorically such as ``to evaluate the models fairly'', and 3) examining perceptions unrelated to LLMs like focusing on different design features of interfaces and evaluating data pre-processing results for LLMs' training instead of LLMs' generated responses. 

We further applied a backward snowballing approach to identify additional relevant papers on people's perception of LLMs with keywords including bias*, ``social harm*'', stereotype*, stigma*, fair*, norm*, ethic*, and safet* in the full papers through the references provided in the introduction and related work sections by the ten eligible papers and meeting the inclusion criteria. Snowballing refers to using the reference lists of a paper (backward) or the citations to the paper (forward) to identify additional papers \cite{wohlin2014guidelines}. As recommended, snowballing could be applied in addition to the search in the databases for a full systematic review \cite{keele2007guidelines, jalali2012systematic}. Using the backward snowballing approach, five more papers were included through the references of the ten eligible papers, resulting in a total number of 15 papers in the final set.


\subsection{Data Extraction and Analysis}
To answer RQ1 and RQ2, we used shared spreadsheets to extract data from the selected papers based on a top-down coding framework. Specifically, for the definitions of biases (RQ1), we extracted the sentences where researchers mentioned the keywords related to bias: {\itshape bias, ``social harm'', stereotype, stigma, fair, norm, ethic, safety} and investigated how these keywords were used by researchers in the context of the included papers. We thematized the bias-related terms based on how the term was used in the included papers to understand what bias-related terms have been investigated by researchers. We further grouped the biases into different themes regarding the sources such as from LLMs or from humans, and biases and related concepts mentioned by researchers in their papers such as bias in datasets or bias of algorithms. For LLM applications (RQ2), we conducted an analysis of the included papers and employed a standardized framework to extract data in order to comprehend the following aspects: 1) the application areas of LLMs, 2) the role of LLMs, 3) algorithms of LLMs, 4) the targeted end-users directly involved in LLMs applications, and 5) the participants recruited for evaluations of LLMs. In order to gain a more detailed understanding of the participants, we also extracted the demographic information of those participants involved in the studies, as presented in the included papers. 

For RQ3, we conducted a content analysis through open coding to extract people's perceptions and applied affinity diagrams to analyze the codes and generate the themes of people's perceptions of LLMs. Guided by grounded theory \cite{strauss1998basics}, two researchers coded people's perceptions within the result sections of included papers using Nvivo software \cite{di2000using}. Each of the generated codes was then transcribed onto a separate digital sticky note on Miro \cite{Miro2023}, a visual workspace platform for teams. To analyze the codes, the research group used an affinity diagram, a technique to organize ideas and issues into patterns and relationships \cite{salminen2022use}, by continually comparing the transcribed sticky notes, integrating them with other notes, and organizing them based on their thematic connections such as similarities, differences, and hierarchical relationships.

\section{FINDINGS}
Following our search process and queries, we found that only a few studies investigated people's perceptions toward bias and related concepts in various applications of LLM. In this review, we only got 15 papers through both database searches and the backward snowballing approach. 
Below, we present the definitions and examples of how the terms related to bias were used by the researchers, the different LLM application areas, and people's perceptions of LLMs reported from these studies.

\subsection{Bias-related Terms and Definitions in Studies About Human Evaluation of LLM}
\label{Sectionofdefinitions}
Of the 15 papers included, only two papers provided an explicit definition of ``bias'' or other related terms: ``For this paper, we will use the term `bias' to refer to any systematic favoring of certain artifacts or behavior over others that are equally valid'' \cite{arnold2018sentiment}, and ``In other cases, humans tend to show automation bias, e.g., automatically relying or over-relying on the output produced by a chatbot.'' \cite{cabrero2023perceived}. The rest of the papers used terms related to bias (i.e., bias*, ``social harm*,'' stereotype*, stigma*, fair*, norm*, ethic*, and safet*) without explicit elaboration on what they might mean. Table \ref{tab:Definition} shows how researchers adopted and used these terms in their papers. Overall, our analysis of the literature shows that researchers investigated three types of biases from two sources: human bias and bias of LLMs. We saw potential opportunities, as well as harm, caused when the two biases interact with each
other.

{\footnotesize\begin{longtable}{p{1.8 cm}p{3 cm}p{9.2 cm}}

\caption{Investigated Biases and Related Concepts in the Included Papers}
\label{tab:Definition}\\
\hline Themes&Biases and related concepts&How the concepts were used in the papers\\ 
\hline
\endfirsthead
\multicolumn{3}{c}%
{\tablename\ \thetable{} -- continued from previous page} \\
\hline Themes&Bias and related concepts&How the concepts were used in the papers\\ 
\hline  
\endhead
\multicolumn{3}{r}{{Continued on next page}} \\
\hline
\endfoot
\hline\hline
\endlastfoot%

    \multirow{4}{*}{\makecell{Individual and \\Social Bias}}&Stereotype&  `` Previous work has shown that different descriptions of gender-based violence (GBV) influence the reader’s perception of who is to blame for the violence, possibly reinforcing \textbf{stereotypes} which see the victim as partly responsible, too.'' \cite{minnema2023responsibility}   \\
    \addlinespace
    \cline{3-3}
    \addlinespace
    
    &Stigma& ``Although therapy can help people practice and learn this Cognitive Reframing of Negative Thoughts, clinician shortages and mental health \textbf{stigma} commonly limit people’s access to therapy.'' \cite{sharma2023cognitive}  \\
    \addlinespace
    \cline{3-3}
    \addlinespace

    &Cultural Norm & ``We expected to find variation between educators and students and across countries in line with different educational approaches and \textbf{cultural norms}. Thus far, we have observed a surprising level of agreement across stakeholders and locals. ''~\cite{smolansky2023educator} ``People generally experience feelings of shame and guilt when they engage in morally unacceptable behaviors or when they violate norms they have internalized''~\cite{chin2020empathy} ``(The low-quality comments) may also be made by peers who intend to be supportive but have difficulties (e.g., being uncertain of the social norms or their own expertise''~\cite{peng2020exploring}\\
    \addlinespace
    \cline{3-3}
    \addlinespace
    
    &Thinking Traps& ``Negative thinking often falls into common patterns, called `thinking traps.' Also called cognitive distortions, these include exaggerated and \textbf{biased patterns of thinking} which cause individuals to perceive reality inaccurately.''~\cite{sharma2023cognitive}
    \\
    \addlinespace
    \hline
    \addlinespace
    \multirow{3}{*}{\makecell{Inaccurate, Unfair,\\ and Biased Nature\\ of LLM Outputs}}
    
    &Datasets containing harmful, biased, toxic content affecting LLM outputs& ``Further, it is well known that commonly used hate-speech datasets are known to have issues with \textbf{bias and fairness}''~\cite{xu2021bot} ``When dialogue models are trained to mimic human-human conversations utilizing large preexisting datasets, they will unfortunately also learn undesirable features from this human-human data, such as the use of \textbf{toxic or biased language}.''~\cite{xu2021bot} ``It is insufficient to merely exclude toxic data from training, as the model would not know how to answer hostile out-of-domain inputs, and \textbf{positive biases} where models tend to agree rather than contradict would lead to undesirable outcomes.''~\cite{xu2021bot} ``We employ 20 annotators to use our designed evaluation tool in order to lessen the \textbf{preference bias of various annotators}.''~\cite{zhang2023glm} ``Like other LLMs, ChatGPT might have intrinsic biases due to imbalanced training data''~\cite{jeblick2022chatgpt}\\
    \addlinespace
    \cline{3-3}
    \addlinespace
    
    & Algorithms affecting the restriction of certain content of the LLM outputs, &``For example, respondents are split in half on the importance of artificial general intelligence, whether language models understand language, and the necessity of linguistic structure and \textbf{inductive bias} for solving NLP problems.''~\cite{michael2022nlp}  ``We also describe training and sampling algorithms that \textbf{bias the generation process} with a specific language style restriction or a topic restriction. ''~\cite{wang2017steering} \\
    \addlinespace
    \cline{3-3}
    \addlinespace

    &Concerns or issues caused by the inaccurate, biased LLM outputs & ``Language models tend to output repetitive and vague responses. They have no model of the truth; they are learning correlations from large amounts of text and thus are able to generate falsehoods. Finally, it has been well-documented that these models can generate offensive language, have \textbf{distributional biases}, and may copy text from the training data.''~\cite{gero2022sparks} ``LLMs may generate text that is semantically plausible and syntactically correct but factually wrong, a phenomenon, known as \textbf{`hallucination'}. The suitability of the term hallucination is questionable as it might imply changes in one’s perceptual experience, which LLMs do not have.''~\cite{kernan2023harnessing}``There are many ways in which NLG tools can become high-risk AI, e.g., producing misleading or inaccurate information, \textbf{biasing the user against a particular social group}, or sharing private information about users.''~\cite{cabrero2023perceived} ``appropriate responses to abusive queries are vital to prevent harmful \textbf{gender biases}.'' \cite{curry2019crowd} \\
    \addlinespace
    \hline
    \addlinespace
    \multirow{3}{*}{\makecell{Opportunistic, \\Sometimes\\ Problematic\\ Influence\\ of LLM\\ Amplified\\ During \\Interactions \\with Human}}
    
    & Spread of bias/misinformation due to confirmation bias and people’s overreliance on the LLM outputs&``Several participants expressed concerns about the potential spread of misinformation and unintentional plagiarism due to the usage of NLG tools... P77 also mentions the risk of \textbf{confirmation bias}.''~\cite{cabrero2023perceived}``In other cases, humans tend to show \textbf{automation bias}, e.g., automatically relying or over-relying on the output produced by a chatbot.''~\cite{cabrero2023perceived}\\
    \addlinespace
    \cline{3-3}
    \addlinespace

    &Unwanted effects of LLMs on humans &``Intervention in high-risk settings such as mental health necessitates ethical considerations related to safety, privacy and \textbf{bias}. There is a possibility that, in attempting to assist, AI may have the \textbf{opposite effect} on people struggling with mental health challenges.''~\cite{sharma2023cognitive}\\
    \addlinespace
    \cline{3-3}
    \addlinespace

    &Inadvertent amplification of these biases due to the vulnerability of humans to biased information& ``We demonstrate that in at least one domain (writing restaurant reviews), \textbf{biased system behavior} leads to biased human behavior: People presented with phrasal text entry shortcuts that were skewed positive wrote more positive reviews than they did when presented with negative-skewed shortcuts.''~\cite{arnold2018sentiment}\\
\end{longtable}}

Human biases and related concepts include stereotypes \cite{minnema2023responsibility}, stigmas \cite{sharma2023cognitive}, negative thinking biases \cite{sharma2023cognitive}, and cultural norms where researchers expect to find variation across countries in line with different practices \cite{curry2019crowd}. Researchers investigated biases and related concepts in terms of models including three aspects: datasets that contain harmful, biased, and toxic content \cite{xu2021bot, zhang2023glm}; algorithms that are integrated with expert inductive bias or tailored according to specific goals such as style transforming, thus affecting the restriction of certain contents of LLM outputs \cite{michael2022nlp, wang2017steering}; and concerns or issues caused by the inaccurate, biased LLM outputs such as distributional biases resulting in too generic responses or ``hallucination''--- where LLMs generate outputs that are semantically plausible and syntactically correct but factually wrong \cite{gero2022sparks, kernan2023harnessing, cabrero2023perceived, curry2019crowd}. When human biases interact with model biases, researchers investigated related concepts about the spread of bias/misinformation due to confirmation bias and people’s overreliance on the LLM outputs \cite{cabrero2023perceived}, the unwanted opposite effects of LLMs on humans such as LLMs developed for positive impacts turning out to bring negative impacts  \cite{sharma2023cognitive}, and the inadvertent amplification and transition of biased behaviors from LLMs to humans due to the vulnerability of humans to biased information \cite{arnold2018sentiment}.


As for the LLMs' role in these biases in the included papers, two papers investigated potential applications and impacts of LLMs in generating positive effects and their help in mitigating such human biases, instead of considering LLMs as the only bias amplifier \cite{minnema2023responsibility, sharma2023cognitive}. Three papers investigate how LLMs can be applied to generate safer conversations when responding to users' abusive behaviors \cite{xu2021bot, chin2020empathy, curry2019crowd}. Two papers examined LLMs as bias producers in providing writing assistance for customer reviews writing and simplifying medical reports, expressing concerns regarding how LLMs could introduce biases \cite{jeblick2022chatgpt, arnold2018sentiment}. The rest papers positioned LLMs in a neutral way and studied LLM applications in different contexts such as applying LLMs in translations, scientific writing, generating knowledge-based dialogue, or generating responses in different personas like Star Wars, Kennedy, Hillary, and Trump bots \cite{cabrero2023perceived, gero2022sparks, kernan2023harnessing, wang2017steering, zhang2023glm, peng2020exploring, michael2022nlp}. 

\subsection{Application Areas of LLMs in Included Studies: Learning About the Role of LLMs, Intended Target Users, and Evaluators of the Studies}


\begin{table}[htbp]
  \caption{Application Areas of LLM in Included Studies}
  \label{tab:application}
  \resizebox{\linewidth}{!}{
  \begin{tabular}{p{0.15\textwidth}p{0.35\textwidth}p{0.15\textwidth}p{0.20\textwidth}p{0.50\textwidth}p{0.07\textwidth}}
    \toprule
    Roles of LLM&Contexts of the Task&Models Used&Target Users&Methods | Human Evaluators &Reference\\
    \midrule
    &LLM rewrites news with different perspectives for gender-based violence news&GPT-3, mBART&Journalists&Quan | 7 females who read the news with different education level within authors' network (Mean age:46) &\cite{minnema2023responsibility}\\
    \addlinespace
    \cline{2-6}
    \addlinespace
    \multirow{4}{*}{\makecell{Content \\transformation}} &LLM performs cognitive reframing of negative thoughts through interactive conversations& GPT-3, RoBERTa&Support seekers&Quan | 3 mental health practitioners; Quan based on a randomized field-study | 2,067 Mental Health America Website visitors &\cite{sharma2023cognitive}\\
    \addlinespace
    \cline{2-6}
    \addlinespace
    &LLM simplifies medical report content & ChatGPT &Patients&Mix | 15 radiologists with varying levels of experience from one clinic &\cite{jeblick2022chatgpt}\\
    \addlinespace
    \cline{2-6}
    \addlinespace
    &LLM influences output style and topic through model training using small `scenting' datasets and decoding methods &Neural Encoder-Decoder & Text-based bot users&Quan | Judges recruited from Amazon Mechanical Turk (AMT) with unknown sample size. Workers were selected based on their AMT prior approval rate (>95\%). Each questionnaire was presented to 3 different workers. &\cite{wang2017steering}\\

    \hline\\

    &LLM generates knowledge-based dialogue with users& GLM-Dialog &Text-based bot users&Quan | 3 annotators for explicit human evaluation and 20 annotators for implicit human evaluation without reported demographic information in the paper &\cite{zhang2023glm}\\
    \addlinespace
    \cline{2-6}
    \addlinespace
    \multirow{3}{*}{Q\&A}  &LLM makes assessments on essays and coding tasks & ChatGPT&Students&Mix | Educators and students from two selective institutions of higher education, one in Australia (338 students; 26 educators) and one in the United States (51 students; 10 educators).&\cite{smolansky2023educator}\\
    \addlinespace
    \cline{2-6}
    \addlinespace
    
    &LLM works as cognitive assistants for troubleshooting e.g., user requesting help with a machine issue from the cognitive assistant& GPT-3.5& Workers&Qual | Initial stage: 3 representatives from a detergent factory and 2 from a textile factory; feature demonstration and focus group stage: 22 developers, and researchers from three different factories(demographic information unavailable) &\cite{kernan2023harnessing}\\

    \hline\\

     &LLM makes writing recommendations for scientific writing& GPT-2; DistilGPT2&Writers&Mix | 13 PhD students from five different STEM disciplines (demographic information unavailable)&\cite{gero2022sparks}\\
    \addlinespace
    \cline{2-6}
    \addlinespace
    \multirow{3}{*}{\makecell{Writing \\assistance}}&LLM suggests next words for writing customer reviews for restaurants&Scalable Modified Kneser-Ney Language Model&Writer composing a restaurant review&Mix based on a within-subjects experiment | 38 students from a university (Demographic information unavailable)& \cite{arnold2018sentiment}\\
    \addlinespace
    \cline{2-6}
    \addlinespace
    &LLM makes recommendations for people in writing posts in online communities to support peers in mental health & BERT & Supporter providers &Mix based on a randomized experiment | 30 students from a local university via word of mouth, having prior experiences of receiving or providing mental support from or to others and that they are willing to offer support to peers in online communities, (13 Females, 15 Males, 2 Not Available; ages ranging from 20 to 30, Mean = 24.37, SD = 2.72) with inclusion criteria regarding fluency in English, depression, and experience of online communities &\cite{peng2020exploring}\\
    \addlinespace
    \hline\\

    &LLM generates safe utterance against users' abusive behavior & GPT-2, DialoGPT, BST 2.7B&Text-based bot users&Quan | At least 3 distinct crowd workers from a disjoint set were instructed to annotate a single utterance (Information on sample size and demographic information unavailable) &\cite{xu2021bot}\\
    \addlinespace
    \cline{2-6}
    \addlinespace
   \multirow{3}{*}{\makecell{Responses \\ to users' \\abusive\\ behaviors}} & LLMs produces different response styles to influence peoples' behaviors & Alexa, Siri, Google Home, Cortana.&Conversational agent users&Quan | 190 crowd workers (130 men, 60 women) on the FigureEight platform, 62.6\% are under age 44&\cite{curry2019crowd}\\
   \addlinespace
    \cline{2-6}
    \addlinespace
   &LLM produces various responses to impact people's behavior by influencing their emotions& Siri, Bixby, Google Assistant, Cortana&Conversational agent users&Mix based on a mixed factorial experiment | 94 students (54 male; 40 female) with ages ranging from 19 to 31 (M=22.78, SD=2.80)&\cite{chin2020empathy}\\
   \addlinespace
   \hline
   \addlinespace
    \multirow{2}{*}{\makecell{Survey\\ \\ Not participants \\testing specific \\ roles of LLMs}}&LLM helps with translation. Examined how different users (e.g., linguists, engineers) perceive and adopt natural language generation tools and their perception of machine-generated text quality. & Any natural language generation tools such as ChatGPT &General users of LLM&Mix | 77 respondents from three groups: 1) linguists, translators, interpreters, or related, 2) participants whose field of expertise is Artificial Intelligence, Natural Language Processing, Computer Science, Software Engineering, or similar fields, and 3) participants in any other field (not reported). &\cite{cabrero2023perceived}\\
   \addlinespace
    \cline{2-6}
    \addlinespace
     &No specific contexts. Examined actively debated issues about LLM (e.g., efficacy, ethics, and intellectual abilities of language models) & Any LLMs& NLP researchers& Quan | 327 researchers co-authored at least 2 ACL publications between 2019 and 2022; Academia 73\%, Industry 22\%, Non-profit 4\%; Senior/faculty 41\%, Junior/postdoc 23\%, PhD student 33\%, Masters student 2\%, Undergraduate 1\%, NA 1\%; United States 58\%, Europe 23\%, Asia/Pacific 8\%, Middle East/North Africa 5\%, Canada 2\%, South America/Caribbean 1\%, Sub-Saharan Africa 0.3\%, Prefer not to say 2\%; Man 67\%, Woman 25\%, Non-binary 3\%, Other/Prefer not to say 6\%; Underrepresented Minority, Yes 26\%, No 63\%, NA 11\% &\cite{michael2022nlp}\\
  \bottomrule
\end{tabular}}
\raggedright\footnotesize{Quan: Quantitative methods; Qual: Qualitative methods; Mix: Mixed Methods.}
\end{table}

In this section, we describe the application areas of LLMs in the included studies. We further examine the application areas of LLMs, the role of LLMs, the algorithms of LLMs, the targeted end-users directly involved in LLM applications, and the participants recruited for evaluations of LLMs. Table \ref{tab:application} shows the summary of the results.

Included studies investigated four broader application areas of LLMs--i.e., content transformation, Question \& Answering (Q\&A), writing assistance, and responses to users' abusive behaviors---across various fields such as journalism, healthcare \& medicine, education, manufacturing, science writing, and customer reviews. For instance, researchers evaluated with human evaluators how LLMs could apply therapeutic techniques to reframe people's cognitive processes for mental health support \cite{sharma2023cognitive}. In another study, LLM rewrote news about gender violence to influence readers' perspectives on who should take responsibility for the violence \cite{minnema2023responsibility}. Researchers also investigated if LLMs can help simplify medical reports for patients \cite{jeblick2022chatgpt}, or change styles and topics according to different personas for the conversations on different topics \cite{wang2017steering}. Researchers also investigated how LLMs can conduct Q\&A for knowledge-based dialogue~\cite{zhang2023glm}, help students complete essays and coding \cite{smolansky2023educator}, and serve as cognitive assistants for knowledge retrieval and troubleshooting and problem-solving \cite{kernan2023harnessing}. 
To provide writing assistance, studies examined applying LLMs to provide recommendations for science writing \cite{gero2022sparks}, writing customer reviews of restaurants \cite{arnold2018sentiment}, and writing online posts to provide mental health support \cite{peng2020exploring}. To reduce users' abusive behaviors, researchers investigated the impacts of different styles of responses on humans, such as empathy style, avoidance style, and counterattacking style \cite{curry2019crowd, chin2020empathy}, and aimed to generate safe conversations through human-in-the-loop methodology to evaluate the responses~\cite{xu2021bot}.

The models used to test these LLMs include GPT-based models such as GPT-2 \cite{radford2019language}, GPT-3 \cite{brown2020language}, GPT-3.5 \cite{GPT3.5}, ChatGPT \cite{chatGPT}, and DialoGPT \cite{zhang2019dialogpt}. Other LLMs included mBART \cite{tang2021multilingual}, roBERTa \cite{barbieri2020tweeteval}, GLM-Dialog \cite{zhang2023glm}, BERT \cite{devlin2019bert}, BST 2.7B \cite{roller2020recipes}, Kneser-Ney language model \cite{heafield2013scalable}, and neural encoder-decoder model \cite{wang2017steering}. LLM applications, such as Alexa, Siri, Cortana, and Google assistants, were also used in the studies. 

Target users of these LLM applications included journalists, patients, students, writers, mental health support seekers and providers, text-based bot users, and conversational agent users as Table \ref{tab:application} shows. In evaluating LLMs, crowd workers and students from universities were the most frequently tested samples. Some studies did not provide details of the demographic statistics of participants and the sample sizes.



\subsection{People's Perceptions of LLMs}
We found four salient themes around how the reviewed papers studied people's perception of LLMs about 1) performances and 2) its anthropomorphism. We also found 3) factors that influence people's perceptions and expectations and 4) concerns for LLMs. Appendix \ref{appendix:Affinitydiagram} introduces the structure of the affinity diagram and the organization of our findings.

\subsubsection{Toward LLM's Performances}
The papers shared how people perceived its advantages and biases. However, these perceptions contradicted one another, depending on the tasks, domains, topics, and individual factors. 

\textbf{Perceived Advantages}. LLMs were perceived as a tool that saves time and gives access to the most up-to-date information quickly.
Furthermore, participants revealed that LLMs were robust enough to respond to poorly phrased questions, unlike other information tools (e.g., Google). 
LLM could also enhance peoples' confidence in writing and communication across cultures while providing compatibility with various user interfaces. 

Participants in studies that used LLMs for translations and posting in online communities for peer support perceived that LLMs could help them save time by efficiently completing tedious tasks \cite{cabrero2023perceived, arnold2018sentiment}. 
Depending on the topic, participants felt that LLMs could offer more accurate information than other internet-based information tools (e.g., Wikipedia). 
For example, participants in a study applying LLMs for scientific writing discovered that LLM-generated responses were more up-to-date on modern psychology topics when compared to those on Wikipedia~\cite{gero2022sparks}. For robustness, factory representatives (e.g., from detergent factories and textile factories) in a study were satisfied with the system's ability to interpret some of their poorly phrased questions \cite{kernan2023harnessing}. Participants in the study that used LLMs to provide suggestions in post-writing to support their peers in online communities stated that LLMs could enhance their confidence and satisfaction in writing \cite{peng2020exploring}. Moreover, some participants highlighted the ease of cross-cultural communication, especially when there is an obvious language barrier \cite{cabrero2023perceived}. Some participants mentioned that they were able to concentrate better on scientific writing tasks with LLM-integrated user interfaces than using web search. For instance, a participant noted that even though their generations and interactions with the LLM were not as good as Google in terms of accuracy, they were able to maintain focus on their own writing without being distracted by subsequent web searches \cite{gero2022sparks}.    

\textbf{Perceived Bias in LLMs}. The studies reported that there was a lack of understanding and awareness among the participants on how bias affects LLM results, although a few participants were aware that inherent biases of people can influence AI models \cite{cabrero2023perceived}. 

People perceived distribution bias as a disadvantage since it led to a lack of diversity and generated overly generic responses. For example, when using LLMs for scientific writing, most participants found that generated outputs were ``less diverse than a Google search'' \cite{gero2022sparks}. Moreover, participants in the study that used LLMs for writing restaurant reviews reported that the lack of specificity in the generated responses made it challenging to integrate them into their own writing \cite{arnold2018sentiment}. The quality of responses could depend on how people prompt LLMs. 

Although most of the participants agreed that machine-generated texts are, in fact, gender-biased---however, half of them stated that even though natural language generation is inherently gender biased, it is as good as humans, or better, in avoiding gender bias \cite{cabrero2023perceived}. 

People tended to perceive biases when they felt out of control of the system and failed to achieve the desired outcomes for their given tasks. 
For example, some participants in the study that used LLMs for scientific writing reported they eventually gave up after trying different prompts multiple times because they ``could not get the prompts to give me that spark [I wanted]''~\cite{gero2022sparks}. 
On the other hand, some participants from another study that used LLMs for writing restaurant reviews stated that, at times, they felt like the words predicted by LLMs were guiding their own writing \cite{arnold2018sentiment}. 
Several users reportedly felt less in control when using LLMs to complete specific tasks, whereas they felt fully in control when using Google for idea generation \cite{gero2022sparks}. 
These experiences might have been a result of ``hallucination'' and/or biased system behaviors, but participants did not report them as biases of LLMs.

\textbf{Conflicting LLM Performances} 
LLM's performances can be measured through task- and domain-dependent measures, such as accuracy, coherence, and impacts, or subjective measures, such as appropriateness, preference, efficiency, effectiveness, and explainability. The results of these measures have varied, often conflicting across the studies.

For task- and domain-dependent measures, variances in LLMs performances across tasks and domains were attributed to the limitations and shortcomings of LLMs.
For example, LLMs performed well for Q\&A \cite{kernan2023harnessing} and translation but were not performing well for simplifying medical reports \cite{jeblick2022chatgpt}. 
Also, factory workers applied LLMs as cognitive assistants for troubleshooting and were impressed by the accuracy of the responses \cite{kernan2023harnessing}. 
When asked to complete translation tasks, one-third (33\%) of a study's participants reported that AI-generated text accuracy was equally good--if not better--than human-written text. However, 19\% of participants expressed uncertainty regarding potential disparities in machine translation performance, especially concerning certain languages, language pairs, or specific domains \cite{cabrero2023perceived}. 
However, in medical contexts, radiologists highlighted encountering inaccurate passages. They also identified missing relevant information as well as potentially harmful conclusions, including misinterpretation of medical terms, imprecise language,  hallucinations, odd language, missing correct information about crucial patients' medical information, and missing key medical findings within the simplified medical reports generated \cite{jeblick2022chatgpt}. 

The perceived coherence of LLMs also showed variation across topics, tasks, and different prompts. Participants using LLMs for scientific writing reported that the coherence of these models is effective for specific subjects but less so for others. For instance, they noted high coherence scores for computer security and green marketing topics, while dynein and automata theory received low coherence scores \cite{gero2022sparks}. Meanwhile, they also stated that certain prompt templates were more effective for specific topics compared to others such as ``One attribute of X is'' prompt templates worked better for source code than old-growth forests topics\cite{gero2022sparks}. 

People's perceived impacts of LLMs also varied among tasks and different groups. For example, students and educators felt that LLMs impacted the assessments differently as essays (including reports, literature reviews, case studies, and research papers), computer code (including pseudo-code and mathematical proofs), short-answer and multiple-choice questions were very or at least moderately impacted by LLMs. The assessments that required product design or creative/artistic work were moderately impacted by LLMs, but other assignment types such as presentations, and discussions that were either pre-recorded or live were not impacted by LLMs \cite{smolansky2023educator}. 


For the subjective measures, such as appropriateness, preference, efficiency, effectiveness, and explainability, the variations in perceived performances of LLMs among individuals may be attributed to personal factors, including demographics within social and cultural contexts. For example, variations were observed in how participants rated the appropriateness of LLMs. When using different commercial conversational agents such as Siri, Cortana, etc. to respond in different styles such as empathy, avoidance, and counterattacking to respond to verbal abuse, most participants preferred the empathy style and enjoyed chatting with the conversational agents while some participants rated counterattacking style as the most appropriate, stating that the ``eye for an eye'' style responses made them regret their actions \cite{chin2020empathy}. Furthermore, Gen Z rated the avoidance strategies adopted by the commercial conversational agents for verbal abuse responses significantly lower than other groups of different ages, whereas older people thought humorous responses to harassment were highly inappropriate \cite{curry2019crowd}. Individual backgrounds could also influence the perception--in the survey study using LLMs for translations, participants with expertise in AI, NLP, Computer Science, Software Engineering, and related fields were generally more positive towards LLMs and their potential applications than participants with expertise in linguistic-related fields and participants with expertise in other fields \cite{cabrero2023perceived}. Educators and students showed different preferences for prompts, collaborative LLMs' responses, and response styles in applying LLMs for essays and coding assessments \cite{smolansky2023educator}.

The perception of LLM's efficiency varied among people as well. Some participants considered LLMs to be efficient by expressing that ``[it] helped me save a lot of time.'' \cite{arnold2018sentiment}, however, some others reported concerns of wasting time, stating: ``but I think I didn't save much time using it, as I was constantly only looking whether the word I was wanting to write appeared in the box'' \cite{arnold2018sentiment}. This holds true in terms of both effectiveness and explainability. For example, when using LLMs in scientific writing, participants' reports of usefulness varied \cite{gero2022sparks}. Perceptions of LLM's explainability differ among people with different levels of expertise. In a survey exploring people's perceptions of LLMs explainability, approximately one-third of respondents believed that ChatGPT clearly communicates its decision-making process to its intended users. Among these participants, those with expertise in AI, NLP, Computer science, software engineering, or similar fields reported a higher percentage of belief in ChatGPT's explainability compared to those in linguistics, while people outside of computing and linguistic communities had the lowest percentage (26\%) of respondents who believed in ChatGPT's explainability \cite{cabrero2023perceived}.

\subsubsection{People's Perceptions of LLM's Perceived Athropomorphism}
The perceived anthropomorphism (i.e., being human-like) of LLMs is a unique performance for technology solutions imitating humans, including likability, perceived intelligence, tone clarity, and distinguishability from humans.  In general, participants had a positive experience with existing commercial conversational agents, stating that these models were interesting, likable, and not boring to interact with \cite{chin2020empathy}. The perceived intelligence and tone clarity were not associated with likability. For example, amongst the different types of commercial conversational agents that responded to verbal abuse, participants rated the counterattacking responses as the same as the empathy style in terms of anthropomorphism and tone clarity \cite{chin2020empathy}. Nowadays, people can still distinguish LLMs from humans in certain tasks \cite{gero2022sparks}. For example, some participants reported seeing distinctions in scientific writing when comparing texts written by humans versus texts generated by LLMs \cite{gero2022sparks}. Moreover, NLP researchers held different opinions on the language comprehension capabilities of LLMS. This division was nearly equal, with half in agreement \cite{michael2022nlp}.

\subsubsection{Factors Influencing People's Perceptions}
We found some hidden factors that may help explain the diversity in people's perceptions of LLMs' performances. These factors include diverse contextual needs, varying expectations of LLMs, and distinct mental models of LLMs. 

People's needs in context showed dynamic patterns. For example, according to 2067 Mental Health American website visitors, they preferred medium-readability, lower-positivity, highly empathic, and highly specific reframes when applying LLMs to reframe negative thoughts for mental health support, showing non-linear patterns across different dimensions. Such needs are challenging to transfer to a mathematical problem with an objective function for models to train since multiple facets of needs should be first investigated.

Furthermore, people had different expectations of what LLMs should or should not do. For example, one participant in the included study investigating LLM applications in science writing stated that LLMs should not take over something that is fundamental to humanity's work \cite{gero2022sparks}. In the survey investigating people's perceptions of LLMs in translation tasks, some participants added that LLMs should be used for good and not for malicious purposes (e.g., generating knowledge and not the detriment of it) \cite{cabrero2023perceived}. 
 
As people increasingly used LLMs and commercial conversational agents, they began to form their own mental models to conceptualize how LLMs would work. For example, some participants thought that LLMs were impartial, and ``assume ChatGPT is an objective entity''~\cite{cabrero2023perceived}, which may explain why people thought LLMs were better than humans at dealing with social biases. In other aspects, people had different mental models of whether LLMs were human-like. In the study investigating different styles of conversational agent responses to abusive behaviors, some participants stated that LLMs as machines ``should not display negative emotions toward the user'' while some other participants preferred that LLMs should be very human-like and it was totally acceptable and interesting at the same time for LLMs to generate counterattacking responses \cite{chin2020empathy}. Some people even expected LLMs to understand the emotional landscape of the user \cite{cabrero2023perceived}.
 
\subsubsection{People's Concerns in LLMs}
Considering the existing perceived performances with hidden factors of people's perceptions, people had concerns about regulation and data protection, and potential negative impacts of LLM applications, but with an overtrust phenomenon in LLMs.

With the emergence of generative AI and commercial LLM-based applications, there were evident concerns regarding usage regulation and data protection. When discussing ethics, and LLM's control of usages, students and educators had ethical concerns about using LLMs--whether or not they should be allowed to use generative AI tools \cite{smolansky2023educator}. Similarly, writers and translators had concerns about unintentional plagiarism and authorship \cite{gero2022sparks}. People applying LLMs in post-writing for peer support raise concerns about the lack of sincerity and authorship \cite{peng2020exploring}. Furthermore, people also had concerns over control and regulation for potential misuse. For example, participants expressed the need for transparency and data protection, with some even stating that LLM-based commercial applications should not be publicly released until ethical issues are mitigated \cite{cabrero2023perceived}.

For data protection, people had concerns about privacy issues when using LLMs for mental health support \cite{peng2020exploring} and using personal information and data \cite{cabrero2023perceived}, the monopoly of companies' singular control of foundational systems such as ChatGPT \cite{michael2022nlp}. Factory representatives expressed concerns related to confidential information and proprietary knowledge being leaked \cite{kernan2023harnessing}. However, opinions on industry influences varied significantly among job sectors, with 82\% of respondents in academia agreeing that private firms have too much influence, compared to only 58\% of respondents in industry \cite{michael2022nlp}.

People also had concerns about LLMs in terms of potential negative impacts, including lack of fact-checking, unintended harms, and potential decline in people's capabilities (i.e., compromised work quality and job loss). In terms of fact-checking, participants overtrusted LLMs because they believed them to be impartial, ``unfiltered information being taken as truthful'', and ``forgetting how to gather information or recognize misinformation due to frequent use of ChatGPT'' \cite{cabrero2023perceived}. However, overtrust issues were influenced by people's backgrounds and experiences. Participants with more experience in information and communications technology reported lower levels of trust in natural language generation tools compared to those with less experience \cite{cabrero2023perceived}. Furthermore, people's concerns about the potential negative impacts of LLMs extended to topics such as misinformation and positive feedback loops, where ``the bias will amplify'' \cite{cabrero2023perceived}. 

When evaluating potential negative impacts of LLMs, people perceived unintended harms as a significant factor, including topics like safety (LLMs are susceptible to unintended harmful attacks), brand images, and the fact that more needs to be done to address potential risks--and that this gap should be bridged \cite{cabrero2023perceived}. Factory representatives expressed concerns that prompts made by their employees may negatively impact their company's brand image \cite{kernan2023harnessing}. According to crowd workers evaluation, LLMs were susceptible to attack, the safe responses generated by LLMs were no more than 60\% \cite{xu2021bot} and participants expressed concerns that the dimensions assessed by LLMs were insufficient to capture the subtle usage of language when offering peer support for mental health \cite{peng2020exploring}.

During creative writing and other goal-oriented tasks, some students reported concerns about a loss of creativity \cite{smolansky2023educator} and some participants had concerns that the natural language generation tools would result in less critical thinking, memory development, knowledge, and deductive capability, correlation thinking, and losing writing and communication skills \cite{cabrero2023perceived}. They also expressed concerns about a decrease in efforts to study different languages and a ``decrease in creativity and an overall trend towards simplistic, repetitive content'' \cite{cabrero2023perceived}. Furthermore, some participants had concerns about the quality of work being sacrificed for time-saving, and immediate results. For example, some participants worried that natural language generation tool users ``would be sacrificing quality translations for immediate results, making translators lose jobs while having subpar machine translations'' and ``the use of natural language synthesis might devalue the position of experts''. Moreover, participants in a survey study expressed significant concern about job losses, particularly the potential for LLMs to replace human translators. These concerns are rooted in the growing automation trends since LLMs are increasingly occupying roles that were once held by humans in the business sector \cite{cabrero2023perceived}.

\section{DISCUSSION}
\subsection{Gaps in Studying Human Experiences and Perceptions Toward Biases of LLM}
Biases have been defined and studied in many contexts. For instance, Ferrara reviewed the source of biases for LLMs, including training data, algorithms, labeling and annotation, product design decisions, and policy decisions \cite{ferrara2023should}. For instance, algorithms can assign greater significance to specific features or data points \cite{ferrara2023should}, or the subjective preferences of human annotators can bias the training data, favoring particular use cases or tailoring user interfaces for specific demographics or industries \cite{ferrara2023should}, and developers could establish policies to either restrict or promote specific model behaviors \cite{ferrara2023should}. Blodgett et al., also similarly described how the biases could stem from the source material or selection process for training data \cite{blodgett2020language}. 

At the same time, in a survey of 146 papers examining ``bias'' in NLP systems, findings showed the research motivations were frequently unclear, inconsistent, and lacking. Additionally, they pointed out a deficiency in the proper use of quantitative techniques for assessing or mitigating ``bias''; and they have limited engagement with relevant literature beyond the scope of NLP \cite{blodgett2020language}. Our results showed similar challenges. The concept of bias and related terms was ill-defined, where, out of 15 included papers, only two papers provided the definitions of bias and related terms in their papers~\cite{arnold2018sentiment, cabrero2023perceived}. Papers used terms related to bias (i.e., bias, social harm, stereotype, stigma, fair, norm, ethic, and safety) without explicitly defining them. Borrowing the recommendations Blodgett et al. suggested to the NLP community, researchers should be able to articulate their conceptualizations of ``bias'' or any other related concepts in terms of ``what kinds of system behaviors are harmful, in what ways, to whom, and why, as well as the normative reasoning underlying these statements'' \cite{blodgett2020language}.

Because of the lack of articulation on what is considered bias, or other related terms, there is a lack of standardized measurements of these concepts. The studies in this review investigated different LLM performances such as coherence, efficiency, accuracy, appropriateness, likability, efficiency, usefulness, etc. For people's controversial perceptions we discovered, that some performances were task-and-domain-dependent such as accuracy, coherence, and LLM impacts on tasks, while some performances were individual-dependent such as likability, appropriateness, and efficiency. These measurements need to be investigated in conjunction with researchers' articulation of the definitions of bias while considering individual factors and application context to establish how these measures can contribute to understanding bias. Although task-and-domain-dependent performance can be measured through metrics and ground truth (i.e., accuracy), for individual-dependent performances, researchers should collect large enough samples of diverse participants regarding demographic distribution. Also, rich qualitative data about individual contexts and reasoning behind perceptions can ensure the successful conceptualization, measurement, and evaluation of biases.


 
Most studies had limited sample sizes and failed to gather demographic information. This finding on the lack of evidence on the involvement of human subjects in LLM research shows that we are still at an early stage of testing people's perceptions and experiences around LLM. Furthermore, a lack of data on who the human evaluators were shows our limited focus on social-cultural factors that are critical to how people form perceptions.
\subsection{Dynamic Perceptions of LLMs due to Imperfect LLM Performance and Hidden Factors}
We found multiple factors influence people's perceptions. Contexts, tasks, users' needs, expectations, and mental models to characterize LLMs will influence people's perceptions of LLMs. However, subjective people’s perceptions of usefulness, efficiency, effectiveness, appropriateness, and preferences of LLMs require more investigations of user experience in interacting with LLMs. There are characteristics some people like while others don't, such as the extent of anthropomorphism, since people have different opinions of how human-like LLMs should be. For instance, regarding the perceptions of the appropriateness of LLMs in generating abusive behavior responses, some participants thought it was appropriate to apply counterattacking styles to respond to abusive behaviors since it is natural for humans to adopt the ``eye for an eye'' tactic while some participants thought it was inappropriate and LLMs should always be nice to people. 

Such a variety of users' needs, expectations, and mental models will influence how people perceive and interact with technology solutions, impacting user experience. Researchers studied people's mental models of AI in a cooperative word-guessing game and found that those who won more often had better estimates of the AI agent's abilities \cite{gero2020mental}. For gestural interactions with spherical displays, researchers found that children and adults had different mental models in the way they verbalized their perceptions about collaborating around the sphere and the physical affordances of the spherical form factor strongly influenced the way both children and adults conceptualized interaction \cite{soni2020adults}. Based on the included studies, people conceptualized LLMs as an alternative to Google search \cite{gero2022sparks} or Wikipedia \cite{gero2022sparks}, resulting in a problematic mental model considering the ``hallucination'' phenomenon. With the integration of LLMs into different interfaces such as keyboards and mobile UIs, further adaptions or transitions of users' mental models are required in the design process. To deliver transparency of LLMs, education could be delivered to users through engaging ways, such as teaching users about each facet of machine learning through the medium of art. This approach hypothesizes that addressing the challenges related to explainability and accountability in AI systems can be facilitated through art and tangible experiences that bridge the gap between computer code and human comprehension \cite{hemment2023ai}.  

\subsection{Implications for CHI}
LLMs gain growing interest in the CHI community. Researchers investigated how prompt engineering can benefit users during the Human-LLM interaction \cite{wu2022ai}. 
Furthermore, Wang et al. investigated adapting LLMs to mobile User Interfaces (UIs), bridging the gap in NLP and graphical UIs \cite{wang2023enabling}. Other LLM research at CHI includes a variety of application areas such as writing \cite{petridis2023anglekindling, dang2023choice}, coding \cite{liu2023wants}, supporting public health interventions \cite{jo2023understanding}, and enhancing augmentative and alternative communication \cite{valencia2023less}. 

Despite the potential, successful adoption of these tools would require people's trust in these systems. Researchers define human-centered AI as supporting its human users ``while revealing its underlying values, biases, limitations, and the ethics of its data gathering and algorithms to foster ethical, interactive, and contestable use''~\cite{Capel2023Human}, we are still at an early stage of testing how people perceive and experience bias in LLMs and being able to communicate these limitations and biases to users effectively~\cite{xu2019explainable, liao2020questioning}. Our review found people's concerns toward LLM and the various aspects of its performance, including coherence, appropriateness, and explainability. Given that LLMs are new and still continuously evolving, continued research should be done to further understand how people's perceptions of these systems also evolve together, and how these negative perceptions can be addressed through novel design solutions. For instance, prompt engineering requires a complex understanding of the model, and users met challenges in generating, evaluating, and explaining prompts' effects \cite{zamfirescu2023johnny}, indicating the necessity of understanding users' mental models of how LLMs work and devising solutions to address user perceptions that might generate challenges around over- or under-trust of these systems. 

Meanwhile, such efforts of mitigating the biases stemming from users are meaningful if the biases inherent in LLMs are also actively being addressed--and this line of work is actively being investigated by the researchers (i.e., ACM Conference on Fairness, Accountability, and Transparency )\cite{FAccT}. Our goal here as a CHI community dedicated to understanding people and improving their experiences would be to articulate and further examine the different types of negative effects that LLMs have for people, how people perceive them, and how people's experiences can be improved and ensured with safety.

\section{LIMITATIONS AND FUTURE WORK}
Though we employed a systematic approach to select keywords related to fairness, bias, privacy, and ethics in the context of using LLMs, this method may not have captured all pertinent literature on human perception of using LLM applications. Articles that include different terminology or focus on nuanced aspects of the related topics may have been inadvertently omitted.
Future work can further develop a taxonomy that categorizes diverse human perceptions when using LLM applications in various contexts.

\section{CONCLUSION}
This paper presented a systematic review of the empirical studies on people's perceptions of LLMs. We screened 15 papers from a total of 231 papers, analyzing different themes around the concept of bias and related terms, and how the concepts were articulated and tested by participants in the included papers. By presenting the application areas, roles of LLMs, LLM models, target users of the LLM application, and the evaluators of the LLMs, we provide insights into what we know about people's perceptions of bias toward LLMs. We discuss how people's perceptions of LLMs are influenced by context factors such as tasks, domains, user needs, expectations, and mental models, which are often contradictory. However, much of the work is missing detailed information on who the evaluators were, leaving gaps in defining bias and related concepts, measuring those concepts, and understanding individual factors that might affect these perceptions. More empirical work with humans and taxonomy and methodologies for user interaction with LLMs considering biases are needed for future research. 


\bibliographystyle{ACM-Reference-Format}
\bibliography{references}

\appendix

\section{Research Methods}

\subsection{Preliminary Work for ACL Anthology Collection in this Literature Review}
\label{appendix:a}
We conducted an initial collection on July 19th, 2023 and retrieved the full Anthology with abstracts as BibTex and exported the data to Excel through Zotero \cite{mueen2011zotero}. Then, we selected long papers, short papers, and industry papers from 73 ACL events held since 2017 spanning 13 long-standing ACL venues.

The 13 venues were the Asian Chapter of the Association for Computational Linguistics (AACL), the Annual Meeting of the Association for Computational Linguistics (ACL), Computational Linguistics (CL), Conference on Computational Natural Language Learning (CoNLL), European Chapter of the Association for Computational Linguistics (EACL), Conference on Empirical Methods in Natural Language Processing (EMNLP), Findings of the Association for Computational Linguistics (Findings), International Conference on Spoken Language Translation (IWSLT), North American Chapter of the Association for Computational Linguistics (NAACL), Lexical and Computational Semantics and Semantic Evaluation (formerly Workshop on Sense Evaluation) (SemEval), Joint Conference on Lexical and Computational Semantics (*SEM), Transactions of the Association for Computational Linguistics (TACL), and Workshop on Statistical Machine Translation (WMT). 

This process resulted in 18,515 papers, among which 574 papers didn't have abstracts. As a result, 17,941 papers remained from ACL Anthology for further query-based search.

\subsection{Affinity Diagram}
\label{appendix:Affinitydiagram}
\begin{figure}[htbp]
  \centering
  \includegraphics[width=\linewidth]{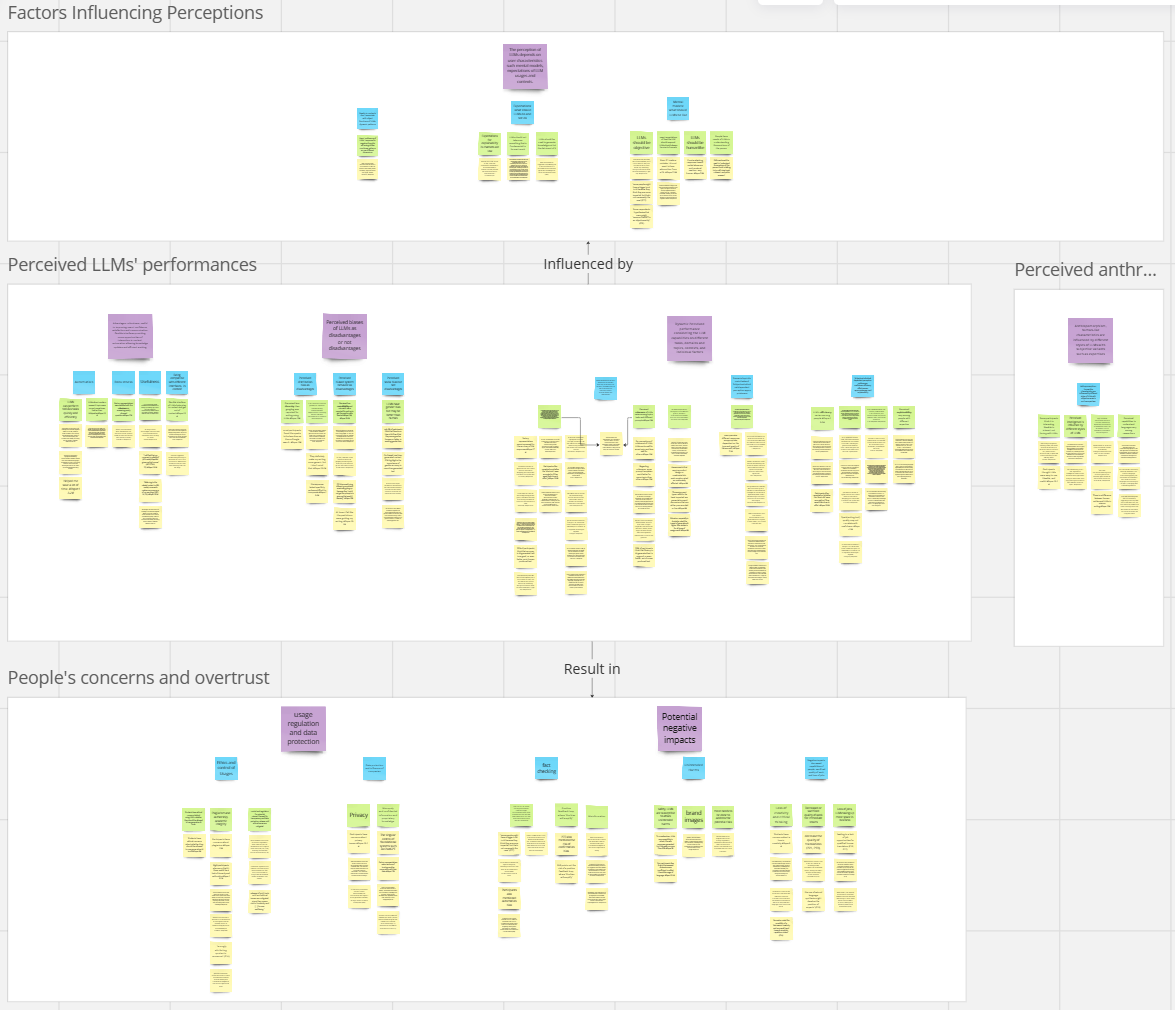}
  \caption{Diagram showing the affinity diagram of people's perception of LLMs.}
  \Description{This figure shows the affinity diagram of people's perception of LLMs, consisting of people's perceptions of LLMs, hidden factors influencing people's perceptions, and people's concerns about LLMs.}
  \label{fig:Affinity}
\end{figure}

The affinity diagram is shown in Figure \ref{fig:Affinity}. The leaf sticky notes in yellow are transcribed codes. The top green sticky notes are topics of the leaf notes. The blue notes are the top codes of the green sticky notes. The largest purple sticky notes are the main themes. Lines refer to the relationships between the linked objects. The arrows refer to the cause-and-effect relations. This affinity diagram was organized through the bottom-up approach.

\end{document}